# Demonstration of input-to-output gain and temporal noise mitigation in a Talbot amplifier

Reza Maram, Mohamed Seghilani, Jinwoo Jeon, Xiao-Zhou Li, Luis Romero Cortés, James van Howe and José Azaña

*Abstract*—: We experimentally demonstrate intensity amplification of repetitive picosecond optical pulses with an input-to-output gain up to 5.5 dB using a passive Talbot amplifier. Through the dispersion-induced temporal Talbot effect, the amplifier uses electro-optic phase modulation and a low-loss dispersive medium to exploit and coherently redistribute the energy of the original pulse train into fewer, replica, amplified pulses. In addition, we show how our passive amplifier mitigates the pulse train timing-jitter and pulse-to-pulse intensity fluctuations above a specific cutoff frequency.

*Index Terms*— Optical pulse train, passive amplifier, temporal Talbot effect, temporal phase modulation, pulse-to-pulse intensity fluctuations, timing jitter, temporal noise mitigation.

## I. Introduction

Intensity amplification of waveforms is necessary for increasing the peak power of signals for initiating physical processes, extracting information from the environment, and communications. In particular, amplifying optical pulses to high intensities is essential for numerous applications, such as generation of high-order harmonics, frequency combs, nonlinear optics, and spectroscopy [1]. Traditional active amplifiers, such as erbium-doped fiber amplifiers (EDFA) and Raman amplifiers, provide a simple, robust platform to increase the intensity of an input signal to a desired level through an active gain process; i.e., they employ an external power source to multiply the incoming signal carrier count [1]. In addition to amplifying the signal, however, this process introduces various forms of noise and distortion, such as amplified spontaneous emission noise (ASE), pulse-to-pulse intensity fluctuations, timing-jitter, and other signal distortions induced by limited-gain frequency bandwidth. In contrast, passive amplification techniques for pulse intensity amplification have been demonstrated based on coherent addition of repetitive pulses [2-4] showing potential immunity to active-gain distortions.

In the past, it was shown that identical copies of a pulse train could coherently add to one another in an external resonator to redistribute the original signal energy into fewer, replica, amplified pulses. In a lossless system, the remaining pulse copies are strengthened by the amount of repetition rate reduction. For example, [2] and [3] showed coherent addition of pulses by storing and retiming them at a lower repetition rate using a high-finesse cavity. They showed impressive passive amplification factors of more than 100. However, such a method requires complex, ultra-precise active phase control of the input signal envelope, and is vulnerable to timing-jitter noise present on the input pulses.

Recently, we presented a proof-of-concept of a simple, all-fiber, passive, intensity amplification technique for picosecond pulses generated from a commercial laser source without phase-stabilization [4]. Rather than resynchronizing pulses in the time-domain, our method uses temporal and spectral phase filtering to reshape the incoming temporal pulse Train. Particularly, our technique relies on dispersion-induced temporal Talbot (self-imaging) effects to precisely redistribute the original pulses' energy into fewer, amplified and undistorted pulses. Similar to previous approaches, the passive gain factor is ideally determined by the amount of repetition rate reduction. Moreover, leveraging the benefits from the standard Talbot effect, this technique also has the ability to mitigate temporal noise present on the input signals, such as reduction of pulse-to-pulse intensity fluctuations and timing jitter [5]. However, large dispersive losses incurred in this initial demonstration, owing to ~120 km of optical fiber required to impose the correct spectral phase, prevented overall input-to-output intensity gain.

In this letter, we experimentally demonstrate a Talbot amplifier with 5.5 dB overall input-to-output gain using a linearly-chirped fiber Bragg grating (LC-FBG) to impose dispersive delay while avoiding large propagation loss. In addition, we experimentally show and evaluate the predicted noise mitigation performance of the Talbot amplifier, confirming how the amplifier can significantly reduce pulse-to-pulse intensity fluctuations and timing-jitter, so long as the fluctuations are above a specific cutoff-frequency.

## II. Principle of Operation

Figure 1 illustrates the concept of our passive amplification technique. This is based on the temporal Talbot or self-imaging phenomena that are observed as a periodic waveform (e.g., pulse) train propagates through a second-order dispersive medium [4, 5]. As shown in Fig. 1, top, in the standard temporal Talbot effect, a flat-phase repetitive input waveform (signal at $z = 0$) is self-imaged without any distortion in its original temporal shape after dispersive propagation through a distance $z_T$ or any integer multiple of this distance (integer Talbot effect).

Manuscript received xxx xx, xxxx; revised xxx xx, xxxx; accepted xxxx xx, xxxx. Date of publication xxx. This work was supported in part by the Natural Science and Engineering Research Council of Canada and Fonds de recherché Nature et technologies. R. Maram, M. Seghilani, J. Jeon, X.-Z. Li, L. Romero Cortés, and Jose Azaña are with the Institut National de la Recherche Scientifique—Énergie, Matériaux et Télécommunications, Montréal QC H5A 1K6, Canada. James Van Howe is with the Department of Physics and Astronomy, Augustana College, Rock Island, Illinois 61201, USA.

Notice that in this model, a linear increase of the amount of dispersion (group-delay slope) is assumed, such as in most practical cases. There also exists an infinite amount of fractional distances, defined by the so-called Talbot condition, which produce rate-multiplied self-images of the original waveform train with correspondingly reduced individual waveform intensity–see examples at the fractional Talbot distances $z_T/2$ and $2z_T/3$, corresponding to increased repetition rates by 2 and 3, Fig. 1, top. In an integer self-image, the uniform phase profile of the input is restored. However, in the multiplied self-images, such as those observed at distances $z_T/2$ and $2z_T/3$, there exists a waveform-to-waveform residual temporal phase. Such residual phases are deterministic and can be analytically calculated for every self-image in the carpet [5]. By using a multiplied image at a fractional distance as the input instead of the conventional phase-free input at $z = 0$, further dispersive propagation to the distance $z_T$ produces an output with an amplified intensity. For example, Fig. 1, bottom, shows how if we condition a typical flat-phase input to behave like the waveform train at $z_T/2$, by proper temporal phase modulation (dashed black line), subsequent propagation through additional $z_T/2$ of dispersive length will lead to the output shown at $z_T$, i.e, half the repetition rate and twice the pulse intensity of the input train signal. Refs [4] give details of the proper spectral phase profiles (e.g., amount of second-order dispersion) and temporal phase functions that are required for different desired passive amplification factors.

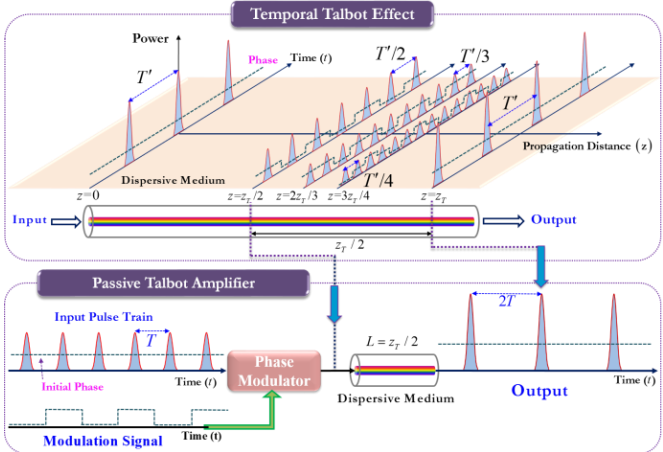

Fig. 1. Passive waveform amplification concept, assuming $m = 3$. $z_T$ is the fundamental Talbot length.

## III. EXPERIMENTAL DEMONSTRATION OF INPUT-TO-OUTPUT GAIN

In a lossless system, because phase-only temporal and spectral manipulation processes ideally preserve the input signal energy, the input-to-output gain should essentially match the Talbot amplification factor. However, insertion losses from practical devices involved in the system counteract the gain from Talbot amplification, and these should be minimized in the design of amplifiers in order to achieve positive input-to-output gain. In [4], we used a large amount of fiber (~120 km of dispersion compensating fiber) with ~42dB insertion losses, which was significantly higher than the largest demonstrated gain factor ($m$=27 or 14.3dB). Therefore, we were unable to show an overall input-to-output gain. However, devices and materials can be engineered to simultaneously provide high levels of dispersion with low insertion loss. For example, here we replace the long fiber with a LC-FBG with a dispersion coefficient $\phi_2 \sim$ 12,930 ps$^2$/rad (~10,000 ps/nm) and ~3 dB loss, to build a device in which system losses are lower than the obtained Talbot amplification gain. In particular, we demonstrate passive amplification with input-to-output gain of 17-ps FWHM optical pulses at a repetition rate of 13.68 GHz generated from a standard commercial pulsed fiber laser. Figs 2(a) and (b) show the prescribed electro-optic phase modulation (EO-PM) profiles applied to the input optical pulses for the cases when we target ideal gain factors of $m = 5$ (=7.0 dB) and 15 (=11.8dB) respectively. The temporal phase functions are calculated from the equations in Ref. [4] and they are experimentally generated from a 7.5 GHz electronic arbitrary waveform generator (AWG), and imposed onto the incoming optical pulse train using a 40-GHz EO-PM.

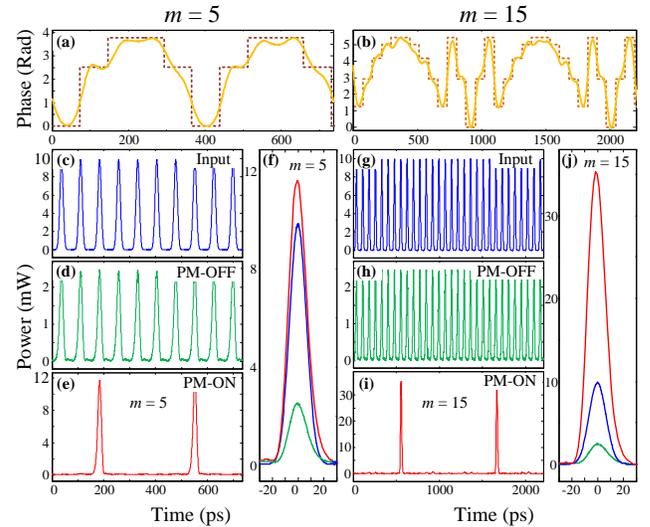

Fig. 2. Experimental results for gain factors of $m = 5$ and 15: (a)-(b) Prescribed temporal phase modulation profiles; ideal phase profiles (dashed brown) and measured phase drives delivered by the AWG (solid orange); (c)-(j) demonstration of passive waveform amplification with input-to-output gain factors of 0.6 and 5.5dB for the cases of $m$=5 and 15, respectively. The optical pulses are measured with a 500-GHz OSO in the averaging mode.

The experimental results of the demonstrated passive amplification are illustrated in Figs. 2(b)-(j). The input signal has a peak power of 9.9 mW (10.0 dBm), shown in Fig. 2(c) and 2(g). The total loss of the system, including the EO-PM and the LC-FBG, is 6.1 dB. In the absence of phase modulation (i.e. when the phase modulator is off, PM-OFF), the output of the LC-FBG is an exact image of the input signal through the integer Talbot condition, with a reduced peak power of 2.5 mW (4.0 dBm) due to the system loss, as shown in Figs. 2(d) and 2(h). Figs. 2(e) and 2(i) show the output pulse trains after the LC-FBG with phase-conditioned input signals (i.e. when the phase modulator is on, PM-ON), for the case of $m$ =5 and 15. The reduced rates after passive amplification are 2.73 and 0.91 GHz respectively, and the corresponding amplified peak powers are 11.6mW (10.6 dBm) and 35.2mW (15.5 dBm), respectively. Figs. 2(f) and 2(j) show the superposition of the input, PM-OFF and PM-ON output individual temporal pulses. The ideal amplification factors ($m$) are nearly obtained when comparing PM-ON and PM-OFF signals, aiming to isolate the effect of the system loss. As shown in Fig. 2(e), the input-to-output gain of the amplifier exceeds slightly (by 0.6 dB) the total loss of the system, in the case of $m = 5$. In the case of $m=15$, shown in Fig 3(e), the input-to-output gain enhances to 5.5 dB. Note that in all cases, the amplified temporal waveforms are nearly undistorted replicas of the input Gaussian pulses.

## IV. ANALYSIS AND EXPERIMENTAL DEMONSTRATION OF TEMPORAL NOISE MITIGATION

Pulse-to-pulse intensity fluctuations and timing jitter are two key factors limiting the performance of any application that relies on the precise regularity and periodicity of optical pulse trains. In the recent past, it has been theoretically analyzed and experimentally demonstrated that the standard temporal Talbot effect has an intrinsic property of mitigating both intensity fluctuations and timing jitter of an optical pulse train [6]. In this section, we present the simulation and experimental results of the intensity-fluctuations and timing-jitter transfer functions for the Talbot amplifier. The intensity fluctuations and timing jitter are modeled respectively by modulating the amplitude and temporal position of the incident 13.65-GHz signal with a sinusoidal wave, using a similar analysis strategy to that employed in previous work [6]. As such, we produce a single noise frequency for timing jitter and pulse-to-pulse fluctuations respectively, and compare the input and output noise levels. The analysis is carried out for different modulation frequencies ($f_M$) to obtain the intensity-fluctuations and timing-jitter transfer functions of the Talbot amplifier.

Fig. 3 shows the intensity-fluctuations transfer function for the frequency range of 30MHz $< f_M <$6GHz, where the incident pulse train optical power is given by

$$P_{in}(t) = \left(1 + A_{in} \sin(2\pi f_M t)\right) \sum_{k=-\infty}^{+\infty} P_0(t - kT) \quad (1)$$

where $P_0(t)$ is the normalized individual Gaussian pulse optical power and $A_{in}$ ($\leq 1$) is the modulating signal's relative amplitude, which determines the maximum offset of the input peak power with respect to the ideal case (uniform pulse train). The intensity-fluctuations transfer function is then defined as the ratio of the maximum offset of the output peak power with respect to $A_{in}$, as a function of the modulation frequency. As such, lower values on the transfer function curve correspond to higher intensity noise mitigation. Our numerical simulations show that the intensity fluctuation in the input pulse train is mitigated more strongly (corresponding to a lower transfer function value) as the modulation frequency is first increased from 0 GHz. From the conducted numerical simulations, we have estimated a cutoff frequency ($f_c$) at which the transfer function reaches down to 0.05 (i.e., remaining 5% of the original amplitude fluctuations) as:

$$f_c \approx 3t_0 / 2\pi \phi_2 \quad (2)$$

where $t_0 = \Delta t/2\sqrt{\ln 2}$ and $\Delta t$ is the FWHM pulse width. Note that the cutoff frequency decreases for shorter pulse widths and/or increased dispersion. The simulation reported here assumes $\Delta t = 17$ ps pulses at a repetition rate of 13.68 GHz. In agreement with the estimate in Eq. (2), Figure 3 shows that the cutoff frequency is nearly similar for all amplification factors, $m=1$ (PM-OFF), 5 and 15 shown here, and is approximately 377 MHz. The estimate in Eq. (2) can be interpreted as follows. First, intensity fluctuations will appear in the spectral domain as noise frequencies between signal tones. For sinusoidal modulation, these are sidebands at a spacing of $f_M$ from signal tones. When the noisy pulse train propagates through dispersive delay, $\phi_2$, pulse energy corresponding to signal tones are properly delayed to form Talbot self-images at integer spacings of the pulse period, $\Delta \tau_G = 2\pi \varphi_2 \Delta f = 2\pi \varphi_2 f_{rep} = NT$ (where $N$ is an integer corresponding to a multiple of the amplification factor). On the other hand, noise frequencies are distributed by dispersion to temporal locations *outside* of the target pulse center. If this delay is *within the pulse width*, noise energy remains as intensity fluctuation distortion. However, if the delay of noise frequencies is large enough, it will distribute outside of the pulse and add to a smaller level at the noise floor between pulses. The overall effect will be to level pulses to the same pulse height, reducing pulse-to-pulse intensity fluctuation. Specifically, when $\Delta \tau_G$ is significantly larger than the pulse width, in particular $3t_0$, i.e., when $\Delta \tau_G = 2\pi \varphi_2 f_m > 3to \Rightarrow f_m > 3to/2\pi \varphi_2$ (corresponding to a cutoff frequency of $f_c = 3t_o/2\pi \varphi_2$), our simulations show that pulse-to-pulse fluctuations are reduced to ~5% of their original value. Moreover, our numerical analysis also shows that the transfer function exhibits resonances at higher modulating frequencies, defined by

$$f_r \approx n (m \times T) / 2\pi \phi_2 \quad (3)$$

where $n$ (= 1, 2, 3, …) denotes the resonance number and $m \times T$ is the output temporal period. These resonances indicate a revival of the original intensity fluctuations in the output pulse train around the corresponding modulating frequencies, as defined by Eq. (3). In our simulated case (see Fig. 3), the first resonance (i.e., $n=1$) appears at $f_r \approx$ 906 MHz for $m = 1$ and at $f_r \approx$ 4.52 GHz for $m = 5$. No resonances are observed for $m = 15$, as the first resonance is to appear at $f_r \approx$ 13.58 GHz, requiring a sinusoidal modulating function being sampled by the 13.68 GHz input pulse train, which would violate the Nyquist condition. We attribute this resonance behavior to the fact that at the resonant frequencies $f_r$ defined by Eq. (3), the relative delay undergone by the corresponding spectral sidebands, $\Delta \tau_{gM}$, becomes equal to the temporal period of the output pulses ($\Delta \tau_{gM} = m \times T$) so that the spectral modulation components of the noise coherently superpose with the temporal pulses. Therefore, the higher the amplification factor $m$, the longer the output temporal period, and this translates into a wider frequency range over which the Talbot amplifier mitigates the input intensity fluctuations. The pass-band of each of these resonances is estimated to be equal to $2 \times f_c$ as well. In view of these results, we conclude that in the practical case of random pulse-to-pulse intensity fluctuations (comprising a broad range of modulating frequencies), a Talbot amplifier will offer an improved noise-mitigation performance as the amplification factor is increased.

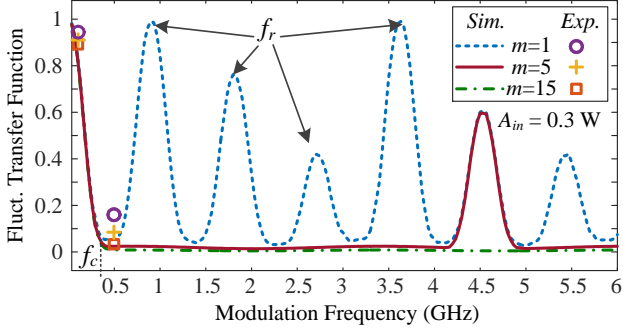

Fig. 3. Intensity-fluctuations transfer function for amplification factors of $m$ = 1, 5, 15. $A_{in}$ is set to 0.3W. The points on the figure show measured experimental data points.

We next experimentally validate the pulse-to-pulse intensity fluctuations mitigation property of the Talbot amplifier. The 17 ps optical pulses generated from the pulsed fiber laser are intensity modulated by a sine wave, generated from a signal generator, using a 40-GHz electro-optic intensity modulator. The measured experimental data points for $m$=1, 5 and 15 are overlaid on the simulation results in Fig. 3 for modulating frequencies below and above the cutoff, namely 50MHz and 500MHz, respectively.

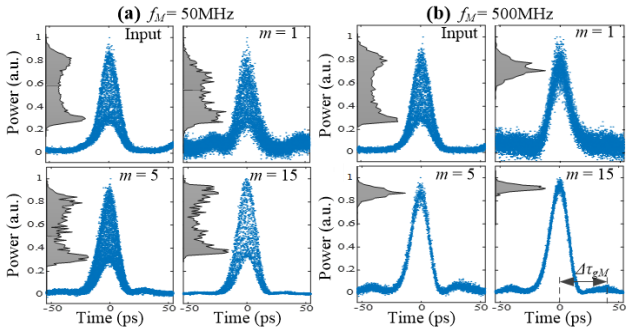

Fig. 4. Measured temporal traces of the Talbot amplifier output for an intensity-modulated input signal with modulation frequencies of (a) 50MHz (below cutoff) and (b) 500MHz (above cutoff).

To calculate the transfer function values, we isolated the effect of EDFA noise and the oscilloscope response by comparing the output results in the presence and absence of the intensity fluctuations in the input pulse train. The corresponding experimental sampling scope traces (persistent mode) are shown in Fig. 4. The results clearly show the expected noise mitigation property of the Talbot amplifier above the cutoff frequency. In Fig. 4(b), the temporal side-lobes can be observed at $\Delta \tau_{gM} = 2\pi \phi_2 \Delta f \approx 40 \mathrm{ps}$, for $\Delta f$=500MHz, away from the output pulses, in agreement with our predictions. Note also the noise distribution conversion from the input (sinusoidal) to the output (Gaussian-like) when the intensity fluctuations are mitigated.

Finally, we evaluate the timing-jitter transfer function for the Talbot amplifier. The incident pulse train is now defined by

$$P_{in}(t) = \sum_{k=-\infty}^{+\infty} P_0\left(t - kT + J_{in} \sin(2\pi f_M t)\right) \quad (4)$$

where $J_{in}$ and $f_M$ are, respectively, the incident timing jitter amplitude, which determines the maximum offset in the pulse's position in time with respect to the ideal case (no jitter), and frequency. The jitter transfer function is then defined as the ratio of the maximum offset in the output pulse's position with respect to $J_{in}$, as a function of the modulating frequency. Fig. 5 depicts the timing jitter transfer function for amplification factors of $m$ = 1, 5 and 15. The 100-fs sampling time used in our simulation limits the recorded minimum value of the transfer function to 0.1. We observe similar trends for the timing jitter transfer function to those described above for the intensity fluctuations, results in Fig 3. Both the cutoff and resonance frequencies of the timing jitter transfer function can be again estimated from Eqs. (2) and (3), respectively, and similar conclusions apply in regard to the timing jitter performance of the Talbot amplifier. Current limitations on equipment prevented us from showing timing-jitter mitigation experimentally.

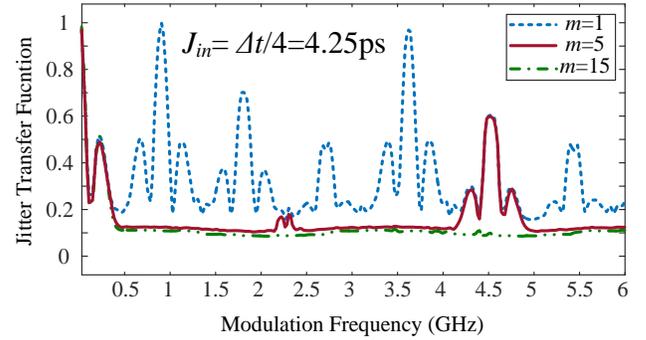

Fig. 5. Jitter transfer function for amplification factors of $m$ = 1, 5, 15.

## V. Conclusion

Through an "inverse" Talbot effect, we have experimentally demonstrated input-to-output intensity gain of picosecond optical pulses up to 5.5 dB using phase modulation combined with linear dispersive propagation in a LC-FBG, without using an active amplification process. In addition, we presented the results of a numerical and experimental investigation of the noise mitigation capabilities of our Talbot amplifier for pulse-to-pulse intensity fluctuations and timing jitter. These results confirm the interest and advantages of the Talbot amplification concept for coherent, linear amplification and regeneration of repetitive waveforms.